\documentstyle[aps,preprint]{revtex}
\preprint{ITP 01-64, PUPT-1994, hep-th/0106242}
\begin{document}
\title{String field theory, non-commutative Chern-Simons theory
and Lie algebra cohomology}
\author{David J. Gross}
\address{Institute for Theoretical Physics,
University of California,
Santa Barbara CA 93106}
\author{Vipul Periwal}
\address{Physics Department,
Princeton University,
Princeton NJ 08544}
\maketitle
\begin{abstract}
Motivated by noncommutative Chern-Simons theory, we construct an infinite
class of field theories that satisfy the axioms of Witten's string field
theory.  These constructions have no propagating open string degrees of
freedom.  We demonstrate the existence of non-trivial classical solutions.
We find  Wilson loop-like observables in these examples.

\end{abstract}
\def\Tr{\hbox{Tr}}
\tightenlines
\section{Introduction}
\medskip
String field theory\cite{sft}\
has enjoyed a renaissance of late\cite{ohm}.  It appears to
provide a framework for concrete calculations to further explore the
ramifications of Sen's ideas on the physical consequences and
the interpretation of tachyon condensation\cite{sen}\ in open string theory.

Some recent work has focussed on string field theory with a modified
BRST operator $Q$ which is pure ghost and linear in ghost
fields\cite{rsz,ell,rsz2,dgt,kaw}.
Such an operator naturally has no
non-trivial physical open string states, and as such provides a putative
explanation for the vanishing of open strings in the stable vacuum of
the bosonic open string theory.  With such a pure ghost $Q$ solutions of
the string field theory equations of motion can be taken to have a 
factorized form
\begin{equation}
\Psi \equiv \Psi_g\otimes\Psi_m: Q\Psi_g + \Psi_g*\Psi_g = 0 \ \hbox{and}\
\Psi_m*\Psi_m = \Psi_m.
\end{equation}
An interesting aspect of these studies is that the BRST operator is linear
in ghost fields, and the integration operation of the string field theory
is left unchanged.  We present in this work an infinite set of
solutions of the axioms of string field theory:
\begin{eqnarray*}
Q^2 &=& 0\\
\int Q\Psi &=& 0 \\
Q(\Psi*\Phi) &=& Q\Psi *\Phi + (-)^{|\Psi|} \Psi * Q\Phi\\
\int \Psi *\Phi &=& (-)^{|\Psi||\Phi|} \int \Phi *\Psi\ \,
\end{eqnarray*}
which have the desired properties:
\begin{enumerate}
\item no physical open string states,
\item closed string observables, and
\item nontrivial classical solutions.
\end{enumerate}
These examples are associated with the cohomology of Lie algebras.  The
only non-obvious aspect of our construction is the definition of an
integration operation on the exterior algebra associated with the
adjoint
representation module of the Lie algebra.

These constructions arise as noncommutative analogues of
Chern-Simons theory\cite{cs}.
Why might this be of interest?  The action of Chern-Simons
theory is formally very similar to the action of Witten's string field
theory, which has been shown to simplify greatly in the limit of strong
noncommutativity.   Chern-Simons theory on commutative manifolds
has correlation functions of Wilson lines as observables---these are
the knot invariants associated with statistical mechanics models on plane
projections of knots in three embedding dimensions.   The natural
observables in
Witten's string field theory seem to be quite different.  From the
perspective of gauge theories on noncommutative spaces, it is actually
difficult to understand what to make of Wilson line expectation values
since such Wilson lines on some examples of  noncommutative manifolds
appear to be gauge invariant only if they are integrated over the manifold.
Obviously, such an integration renders a knot invariant interpretation of
Wilson line correlation functions difficult.  Thus it seems likely that
observables of noncommutative Chern-Simons gauge theory might be much more
interesting from the perspective of models of Witten's string field
theory than commutative Chern-Simons theory.

This paper is organized as follows: In section 2, we consider
the problem of defining an odd-dimensional non-commutative manifold.
We carry this out
explicitly for a three-dimensional case.  In section 3 we
construct the Chern-Simons theory associated with this non-commutative
threefold.  A general framework for finding infinite classes of
examples is then immediate.
In the concluding section, we find nontrivial closed string-like
observables for the
gauge theory, after explaining why there are no open string states in the
spectrum.

\section{Non-commutativity in three dimensions}

What is a natural notion of a odd-dimensional noncommutative manifold?
If we start with a non-commutative even-dimensional manifold associated
with a deformation quantization of a Poisson structure, it would appear
that there are two possible ways in which an odd-dimensional noncommutative
manifold might be defined, either as a contact submanifold associated with
a choice of a level set for a Hamiltonian or as a noncommutative contact
manifold
in one higher dimension associated with a time-dependent Hamiltonian.
We consider the former construction in order to define a noncommutative
threefold.

We shall start with an explicit example and then abstract from the
discussion to arrive at a conclusion that is very simple to state:
Noncommutative Chern-Simons gauge theory is formulated
along the lines of Witten's string field theory with Lie algebra cohomology
playing the r\^ole of $Q.$

If $P$ is a symplectic manifold with a Hamiltonian
function $H$ and $\Sigma(E)$ is a regular energy surface, then the
`restriction' to $\Sigma(E)$ is a contact manifold.  The two-form
$\omega$ is just the pullback of the symplectic form of $P.$
Exact contact manifolds locally have a one-form that
takes the form $\theta = dw + p_{i}
dq^{i},$  where $(w,p_{i},q^{i})$ are the local coordinates.

We can obviously define a noncommutative structure on any
symplectic 4-manifold.  Work with
the simplest 4d symplectic structure $\omega = dp_{1}dq^{1} +
dp_{2}dq^{2}.$  The noncommutative structure defined by this $\omega$
just factorizes into a tensor product of two isomorphic copies of the
algebra associated with the
non-commutative plane---we will denote this algebra by $M.$
Now we define the algebraic analogue of
restricting to $\Sigma(E).$ The first thing we need is to pick an
energy function.  To find a good algebra to associate to constant
energy surfaces in this phase space, we define first
\begin{equation}
I_{E} = \{f *(H-E) |f \in M\}.
\end{equation}
This is the left ideal of functions that vanish on the constant energy surface.
By definition if $g\in I_{E}$ then $h * g \in I_{E}$ by the
associativity of the $*$ product.  We now want to
find a subalgebra $N_{E}$
of $M$ such that $I_{E}$ is a two-sided ideal in $N_{E}.$  It will
then follow that $N_{E}/I_{E}$ is an algebra---$M/I$ is not an
algebra in general unless $I$ is a two-sided ideal in $M.$
Define
\begin{equation}
N_{E}= \{ f\in M| (H-E) * f \in I_{E}\}.
\end{equation}
What this means explicitly is that
for any element $f$ in $N_{E}$ there is an element $g$ in $M$ such
that
\begin{equation}
  (H-E) * f = g * (H-E).
\end{equation}
Thus $I_{E}$ is a two-sided ideal in $N_{E}.$
The quotient algebra ${\cal M}\equiv
N_{E}/I_{E}$ is the noncommutative analogue of
the algebra  of functions on a contact manifold.

For the phase space of two decoupled identical harmonic oscillators 
(with creation and
annihilation operators $a_{i}^{\dagger},a_{i};\  i=1,2,$
and Hamiltonian $H=a_{1}^{\dagger}a_{1} +a_{2}^{\dagger}a_{2}+ 1 $ ), 
for example,  it is
easy to work out what operators are in
$N_{E}.$  These are operators that do not change the total occupation
number of the two decoupled harmonic oscillators.  So the quotient
algebra $\cal M$ is generated by $\zeta \equiv a_{1}^{\dagger}a_{2}, $
$\zeta ^{\dagger}$ and $h\equiv a_{1}^{\dagger}a_{1} -a_{2}^{\dagger}a_{2}.$
These operators satisfy the SU(2) commutation relations, not surprising
since we expect the energy surface to be topologically an $S^3$, 
namely the surface
$H=(x_1^2+p_1^2+x_2^2+p_2^2)/2=E$.
For appropriate values of the energy, the quotient algebra has the well-known
finite-dimensional representations:
\begin{equation}
T_\zeta  = \left(\matrix{0&n&0&\cdots&0\cr
0&0&n-1&\cdots&0\cr
\vdots&\ddots&\ddots&\ddots&\vdots\cr
0&0&\ddots&\ddots&1\cr
0&0&\cdots&0&0\cr}\right),\quad
T_{\zeta^\dagger}=\left(\matrix{0&0&\cdots&0&0\cr
1&0&\cdots&0&0\cr
0&2&\ddots&0&0\cr
\vdots&\ddots&\ddots&\ddots&\vdots\cr
0&0&\cdots&n&0\cr}\right),
\end{equation}
\begin{equation}
T_h = \left(\matrix{n&0&\cdots&0&0\cr
0&n-1&\cdots&0&0\cr
\vdots&\ddots&\ddots&\vdots&\vdots\cr
0&0&\cdots&-n+2&0\cr
0&0&\cdots&0&-n\cr}\right), \quad [T_a,T_b]= f^c_{ab}T_c \, .
\end{equation}
The Casimir element $C=\zeta\zeta^\dagger  +  \zeta^\dagger\zeta + h^2/2$
generates the centre of the enveloping algebra.

Now we turn to the algebra of forms over $\cal M.$  We
introduce   operators $c^a$ and $b_a$ where $a\in
\{\zeta,\zeta^\dagger,h\},$ such that $\{c^a,b_d\} = \delta^a_d$ with
$\{b,b\}=0=\{c,c\}.$
Then the operator
\begin{equation}
Q = c^a T_a - {1\over 2} f_{de}^g c^dc^eb_g
\end{equation}
squares to zero and is the well-known operator computing Lie algebra
cohomology with values in the representation module specified by the
matrices $T.$ Note that $Q^2=0$ for any Lie algebra, the only properties of
$f^a_{bc}$ used in proving this are $f^a_{bc}= f^a_{cb}$ and the Jacobi 
identity.
Also note that
\begin{equation}
\{Q, b_a\} =   T_a -  f_{ae}^g  c^eb_g \equiv {\cal T}_a \, .
\end{equation}
${\cal T}_a$ consists of two pieces, the first  rotates elements of the algebra
in the representation of the $T$'s and the second rotates the ghosts in 
the adjoint
representation.

The interpretation here is  that $Q$ is the noncommutative
exterior derivative, with the algebra of forms given by ${\cal M}[c^a],$ the
(graded-)commutative polynomial
algebra generated by $c^a$ with coefficients in $\cal M.$
With the definition $[b_a, x] = 0 = [c^a,x]$ for any $x\in \cal M$ we see that
$Q$ maps forms to forms, raising the degree by one.
Notice that $\{Q,c^{a}\}+ {1\over 2} f_{bc}^{a}c^{b}c^{c}=0,$
so the $c^{a}$ form an orthonormal basis and ${1\over 2} f_{bc}^{a}c^{b}$
is the spin connection.

\section{Non-commutative Chern-Simons theory}

To define a gauge theory over this
algebra we need to pick a projective $\cal M$-module and define the
Chern-Simons integrand in terms of a connection on this module.  The
simplest example of a module is of course $\cal M$ itself, which defines
for us a U(1) gauge theory.  The integrand is easy enough to write down
formally but it isn't completely obvious what integration is
appropriate.  We define
\begin{equation}
S_{ncACS} = \Tr \int_c \left[\Psi\{Q,\Psi\} + {2\over
3}\Psi^3\right],
\end{equation}
where $\Psi= c^{a}A_{a}$ is a 1-form with components $A_{a}.$
We expect that there should be some natural notion of integration for a
three-form, and we expect that the trace over matrix indices in the
representation $T$ should be the analogue of integration over the threefold.

Recall that the non-Abelian commutative Chern-Simons action takes the form
\begin{equation}
S_{CS}=\int \epsilon^{ijk}\left( \delta_{ab} A^a_i \partial_j A^b_k +
{2\over 3}
f_{abc}  A^a_i A^b_j A^c_k\right)\ ,
\end{equation}
where $A_{i}dx^{i}$ is the gauge potential.
In that context the trace over the adjoint representation provides the
factors $\delta_{ab}$ and $f_{abc}.$  In our case, that of Abelian but
noncommutative gauge theory, the matrix trace takes the place of   $\int
\epsilon^{ijk}.$  We therefore have to figure out what to do with the
cubic term in $c^a$ that takes the place of the three-form.  Given the
constraint that the integration  $\int_c c^ac^bc^d$ must be
completely antisymmetric, the only natural assignment
is to define
\begin{equation}
\int_c c^ac^bc^d = f^{abd}\, ,
\end{equation}
where $f$ are the structure constants of the Lie algebra of SU(2).

Thus far we have restricted to the case of SU(2) which arises naturally
from our contact manifold construction.  It turns out that there is
nothing specific to SU(2) in our
construction, so henceforth we shall work with an arbitrary
semi-simple Lie
algebra ${\cal G}$ and an arbitrary (possibly reducible)
representation $T$ acting on
a vector space $V.$   In particular, ${\cal M}$ will be the complete
operator algebra acting on $V,$ not just the enveloping algebra of
${\cal G}.$  ${\cal M}$ coincides with the enveloping algebra if
and only if $T$ is an irreducible representation.

To verify that this is a suitably gauge-invariant action,   that
the gauge invariance is the usual non-Abelian gauge invariance,
and that the axioms of string field theory are satisfied,  we
first note that the action of $Q$ on a field $\Psi$ of
ghost number $|\Psi|$ is $Q\Psi -(-1)^{|\Psi|}\Psi Q.$
Furthermore,
\begin{equation}
\Tr  \int_c [Q,\chi] = 0
\end{equation}
for any $\chi .$ In fact, only $|\chi|=2 $ can contribute
 so the only choice of $\epsilon$
that can give a non-vanishing contribution given our
definition of $\int_{c}$ is
\begin{equation}
\chi= \chi_{ab}c^{a}c^{b}
\end{equation}
where $\chi_{ab}$ is a matrix in $\cal M$ for every choice of the
indices $a$ and $b.$  Clearly $\chi_{ab}= - \chi_{ba}.$

 The gauge transformations take the form
\begin{equation}
\delta \Psi \equiv [Q + \Psi,\epsilon] \, ,
\end{equation}
 where $\epsilon$ is a ghost number zero field.  Any $\epsilon$ not
satisfying this would of course trivially leave the action invariant.

We can  make this completely explicit in terms of component fields.
It can be directly verified that
\begin{equation}
S_{ncACS} = f^{abc}\Tr \left( A_{a}[T_{b},A_{c}] -{1\over 2}
f^{bcd}A_{a}A_{d}+ {2\over 3}
A_{a}A_{b}A_{c}\right)
\end{equation}
is invariant under
\begin{equation}
\delta_{\epsilon} A_{a} \equiv [{\cal T}_{a} + A_{a},
\epsilon]=[T_{a} + A_{a},\epsilon]\, ,
\end{equation}
for any $\epsilon$ in $\cal M.$
We have therefore arrived at a definition of a non-commutative
Chern-Simons action which satisfies all the axioms of Witten's string
field theory.  Note the  presence of the spin connection coupling in
$S_{ncACS}.$

A natural question is: What does this general noncommutative Abelian
Chern-Simons action have to do with threefolds?  The answer requires
first asking what characteristics one wants in a threefold.
In the
noncommutative setting, an abstract threefold is something that can
be paired with three-forms to give a number.  So the logic is: Given
an algebra $\cal M$ one can define an exterior differential calculus
over the algebra.  The next step involves the analogue of
integrating over a threefold, and that corresponds to finding a
character of this algebra.  This is what the integration operation
given above defines.  It would be interesting to find explicit
energy functions on symplectic manifolds such that the algebras
associated with the non-commutative analogues of
contact  manifolds are explicitly generated by semi-simple Lie
algebra representations.  An extension of our 
construction to the case of superalgebras may also be of interest.  
In that case it would be natural for  $\cal M$ to contain
anticommuting elements as well which would lead to more component
fields in $\Psi.$

\section{Solutions and observables}

We turn now to the issue of classical solutions and
observables\cite{rsz2,dgt}.
\subsection{Classical solutions}
The first observation is that the equation of motion
$(Q+\Psi)^{2}=0$ may be written in components as
\begin{equation}
[T_{b}+A_{b},T_{c}+A_{c}]f^{abc} = f^{abc}f^{dbc}(T_{d}+A_{d}) = C(G)
(T_{a}+A_{a}),
\end{equation}
which essentially defines Lie algebras.
At a solution of this equation, the value of the action is
\begin{equation}
     S_{\rm cl} = {1\over 2} \Tr(A_{a}T_{a}) C(G).
\end{equation}
Are there any nontrivial solutions of this equation of motion?
If $T$ is a finite-dimensional
reducible representation then in fact there is a
very simple set of nontrivial solutions.  If $P$ is the projector
onto a given sub-representation of $T,$ then $A=-PTP$ is a solution
of the classical equation of motion such that
$T+A=(1-P)T=T(1-P).$

Considerably more interesting solutions are possible in the
infinite-dimensional case.  If $S^{\dagger}$ is an operator satisfying
$ S^{\dagger}S=1$ and $SS^{\dagger} = 1-P$ where $P$ is a projection,
we can construct new solutions by setting $T+A=S TS^{\dagger}$ since we
have then
\begin{equation}
S^{\dagger}[T_{b},T_{c}]Sf^{abc} =  C(G)S T_{a}S^{\dagger}.
\end{equation}
$S$ is an example of a partial isometry, so there is a direct
connection to a charge in operator $K$ theory corresponding to
such solutions\cite{wegge}.  Let us compute the value of the
classical action for these solutions.  We find
\begin{equation}
S_{\rm cl} = {1\over 2} \Tr((S T_{a}S^{\dagger}-T_{a})T_{a}) C(G) =
{1\over 2} C(G) [\Tr(S T_{a}S^{\dagger}T_{a})-\Tr(T_{a}T_{a})].
\end{equation}
We cannot, however,  directly deduce that this value is proportional to
the rank of $P$ since the action of the operator $S$  is not so easily
disentangled from the representation $T.$
It is interesting to note that the quantum theory has
off-diagonal fluctuations in $A$ which couple different irreducible
sub-representations in any given solution of these classical
equations, which heuristically are analogous to closed
string couplings generated by open string loops.

\subsection{Physical states}

  In the BRST formalism
we want to compute the cohomology of $Q+\Psi.$  There are several ways
to compute this cohomology, which turns out to be the
cohomology of the Lie algebra, $ {\cal G}.$
Recall the definition of Lie algebra cohomology: If $V$ is the
representation space of a representation $T$ of $\cal G,$ an
$n$-dimensional $V$-cochain is a skew-symmetric multilinear mapping:
${\cal G}^{\times n} \rightarrow V.$  The coboundary operator $s$ is
defined by its action on $n$-cochains:
\begin{eqnarray*}
     (s\omega)(X_{1},\ldots X_{n+1}) &=& \sum_{i=1}^{n+1} (-)^{i+1} T(X_{i})
     \omega(X_{1},\ldots \hat X_{i},\ldots X_{n+1})\\
     &+& \sum_{j,k=1;j<k}^{n+1}(-)^{j+k}\omega([X_{j},X_{k}],
     X_{1},\ldots \hat X_{j},\ldots \hat X_{k},\ldots X_{n+1}),
\end{eqnarray*}
where $X_{i}$ are elements of  $\cal G,$  and $\hat{}$ denotes omission.
For example, $s\omega(X_{1},X_{2}) =
T(X_{1})\omega(X_{2}) -T(X_{2})\omega(X_{1}) -\omega([X_{1},X_{2}).$
Using the fact that $T$ is a representation, it is trivial to verify
that $s^{2}=0.$  Thus cohomology groups for $s$ associated with $T$
with values in $V$ are readily defined as the vector space of cocycles
(cochains annihilated by $s$) modulo the coboundaries (cochains of the
form $s\omega$).  This $s$ operator is
related to $Q$ by a factor of $(n+1)$ for an $n$-cochain
so the cohomology groups of $Q$ and $s$ coincide.
For example,
$H^{0}_{T}({\cal G},V) = V^{\cal G},$ where $V^{\cal G}$ is the set of
vectors left invariant by the action of $\cal G.$ This result is obtained by
noting that  zero-cochains, which are just vectors in $V,$ are
zero-cocycles if and only if  $sv(X) = T(X)v = 0 $ for all $X$ in $\cal G.$
In particular, for irreducible representations there are no invariant
vectors.
It is also  easy to show that for semisimple $\cal G,$
$H^{1}_{T}({\cal G},V)=0:$ A 1-cocycle
is a 1-cochain that satisfies
\begin{equation}
   s\omega(X_{1},X_{2}) =
T(X_{1})\omega(X_{2}) -T(X_{2})\omega(X_{1}) -\omega([X_{1},X_{2}])=0
\end{equation}
for all $X_{1},X_{2}$ in $\cal G.$  On the other hand, a 1-cocycle
$\varpi$
is a coboundary if $\varpi(X) = T(X)v$ for all $X$ in $\cal G$ and
some $v$ in $V.$  Define   linear maps $h(Y)$ (for $Y$ in $\cal G$)
which take  1-cocycles to
1-coboundaries by $(h(Y)\omega)(X) = T(Y)\omega(X) -\omega([Y,X]) =
T(X)\omega(Y).$  Now suppose there is a 1-cocycle $\omega$ which
is annihilated by $h(Y)$ for all $Y$ in $\cal G.$  This means
$T(X)\omega(Y)=0$ for all $X,Y$ in $\cal G,$ but then
using the definition of a 1-cocycle, it follows that $\omega([X,Y])=0$
for all $X,Y$ in $\cal G.$  If $\cal G$ is semisimple, $[\cal G,G]=G,$ so
$\omega$ vanishes on all of $\cal G.$  This is what we wished to
demonstrate.  With a little more algebra one can also show
$H^{2}_{T}({\cal G},V)=0.$

Applying this cohomology computation to our theory, we see that
there are no observables at ghost number $1$ or $2$ for any
representation $T.$ At ghost number $0$ the only observables are
invariants of the representation $T.$
So there would appear to be no nontrivial perturbative open string states
in our theory for semisimple Lie algebras.

\subsection{Observables}

What are the observables in this theory?
The covariant derivative operator
 $${\cal D}=Q+\Psi  $$
 transforms
homogeneously under gauge transformations, 
${\cal D}\to {\cal D}+[{\cal D},\epsilon]$.
In components we can define
\begin{equation}
{\cal D}_a\equiv  \{ b_{a},Q+\Psi\}= {\cal T}_{a}+A_{a} \, ,
\end{equation}
 or separate out the matter component to define
\begin{equation}
  D_a\equiv   T_{a}+A_{a}\, ,
\end{equation}
Both ${\cal D}_a$ and $D_a$ transfrom covariantly under gauge transformations
and therefore any product of the form
\begin{equation}
{\cal O}=\Tr \prod_{i} ({\cal T}_{a_{i}}+A_{a_{i}}), \quad {\rm or} 
\quad O =\Tr
\prod_{i} (T_{a_{i}}+A_{a_{i}})
\end{equation}
is gauge-invariant.   A different basis for this set of observables
is
\begin{equation}
W[\lambda_i] \equiv \Tr \prod_{i} \exp(\lambda_{i}(T_{a_{i}}+A_{a_{i}}) )
\end{equation}
which are like the Wilson lines studied in non-commutative gauge
theories.

We can make this analogy stronger by
recalling equation (22) and noting the anitcommutation with $b_a$
is differentiation with respect to $c_a$.
However, anticommuting differentiation is the same as integration
so we can alternatively think of an abstract set of non-commutative
1-cycles, much as we thought of integration on a non-commutative
threefold, indexed by the generators of $\cal G.$ Define
\begin{equation}
     \oint_{b} c^{a}  = \delta^{a}_{b}.
\end{equation}
Then
\begin{equation}
     \oint_{b} (Q+\Psi)  = {\cal T}_{b} + A_{b}.
\end{equation}
This is written as an operator equation.  We can now write
\begin{equation}
W[\lambda_i]=\Tr \prod_{i} \exp(\lambda_{i}(T_{a_{i}}+A_{a_{i}}) )
=
\Tr \prod_{i} \exp(\lambda_{i} \oint_{a_{i}} (Q+\Psi) )
=
\Tr P\exp( \oint_{\sum \lambda_{i}a_{i}} (Q+\Psi) )
\end{equation}
where $\sum\lambda_{i}a_{i}$ is formally a (path ordered) 1-cycle. 
The ghost part of the
group element factors out.  
Thus in the vacuum ($\Psi=0 $), or more generally in the background 
of a classical solution,
where $(Q+\Psi)^{2}=0$,  $W[\lambda_i]$ is a product of
group elements in the representation associated with
$Q+\Psi $, an element in the loop group. Thus the Wilson
loop, $W[\lambda_i]$, associates with every path in the 
$d$-dimensional space ($d$=
rank of the Lie algebra) given by the ordered sequence 
\{$\lambda_1,\lambda_2, \dots$ \},
the trace of the appropriate group transformation.

When we vary the parameters
$\lambda_{i}$ the variation in the observables amounts to an insertion 
of the commutator
$[T_{a}+A_{a},T_{b}+A_{b}]$ which can be written as an insertion of the
equation of motion, which leads to contact terms similar to
skein relations, but one also obtains an insertion of a term
linear in $T+A$ since there is a torsion term in the curvature.
Lastly, we note that these
expressions are also  reminiscent of Wilson lines in Eguchi-Kawai
reduction, with the index $a$ in the Lie algebra representing the
direction.

These   models may be useful  in the limit of
large representation size  and/or large group rank, as approximations to
string theory, along the lines of the matrix representation of
membrane dynamics.
It may be possible to generalize these models to
$A_{\infty}$ algebras as well.  Defining a cubic action
non-perturbatively is a problem that has not been solved
for commutative Chern-Simons theory, so there is still some work
remaining.

\section{Acknowledgents}
VP is grateful to J. Minahan for several valuable conversations  
during the early stages of this work and E. Witten for some helpful 
comments.  
The work of DJG was supported by the NSF under  the
grants  PHY 99-07949 and PHY 97-22022.  VP was supported by the NSF 
under grant PHY 98-02484.


\begin{thebibliography}{99}
\bibitem{sft} E. Witten, Nucl. Phys. {\bf B268}, 253 (1986)
\bibitem{ohm} K. Ohmori, `A review on tachyon condensation in open
string field theories', hep-th/0102085
\bibitem{sen} A. Sen, JHEP {\bf 9912}, 027 (1999)
\bibitem{rsz} L. Rastelli, A. Sen and B. Zwiebach, `String field
theory around the tachyon vacuum', hep-th/0012251
\bibitem{ell} I. Ellwood and W. Taylor,
`Open string field theory without open strings', hep-th/0103085;
\bibitem{rsz2}L. Rastelli, A. Sen and B. Zwiebach, `Classical solutions 
in string
field theory around the tachyon vacuum', hep-th/0102112; `Half-strings,
projectors and multiple D-branes in vacuum string field theory',
hep-th/0105058;
\bibitem{dgt}D.J. Gross and W. Taylor, `Split string field theory
I,II', hep-th/0105059,0106036
\bibitem{kaw} T. Kawano and K. Okuyama,
`Open string fields as matrices', hep-th/0105129
\bibitem{cs} E. Witten, Comm. Math. Phys. {\bf 121}, 351 (1989)
\bibitem{wegge} N.E. Wegge-Olsen, {\sl K-theory and C*-algebras}, Oxford
University Press, 1993
\end{thebibliography}
\end{document}